\documentstyle[psfig]{texas}

\begin{document}

\title{Photomeson production in astrophysical sources}
\author{A.~M\"ucke$^{1}$, J.P.~Rachen$^{2,3}$, R.~Engel$^4$,
R.J.~Protheroe$^1$ and T.~Stanev$^4$} 
\address{(1) University of Adelaide\\
Dept. of Physics \& Math.Physics, Adelaide, SA 5005, Australia
}
\address{(2) Pennsylvania State University\\
Dept. of Astronomy, University Park, PA 16802, USA}
\address{(3) Universiteit Utrecht\\
Sterrenkundig Instituut, Princetonplein 5, 3584 CC Utrecht, The Netherlands}
\address{(4) Bartol Research Institute\\
University of Delaware, Newark, DE 19716, USA\\
{\rm Email:  amuecke@physics.adelaide.edu.au, J.P.Rachen@astro.uu.nl,\\
eng@lepton.bartol.udel.edu, stanev@bartol.udel.edu,\\ rprother@physics.adelaide.edu.au}}

\begin{abstract}
Photomeson production is the main energy loss for relativistic nucleons in
dense radiation fields like the cosmic microwave background and the radiation
fields in Gamma Ray Bursts (GRB) and jets of Active Galactic Nuclei (AGN).
In this paper we study photomeson production in typical GRB and AGN jet
radiation fields by using the recently developed Monte Carlo event generator
SOPHIA (see these proceedings). We discuss processes that are
relevant for the physics of cosmic ray acceleration and the production
of neutrinos and gamma rays. We compare our results with widely used
approximations, and find significant deviations, particularly for GRBs.
The photoproduction of antibaryons as a so far not considered
effect in astrophysics is briefly discussed.

\end{abstract}

\section{Introduction}
 Ultrahigh-energy nucleons propagating through dense radiation fields
 lose their energy mainly through photomeson production. Radiation fields
 are universal (Cosmic Microwave Background Radiation (CMBR)) and
 are very intense at luminous astrophysical sources such as 
Active Galactic Nuclei (AGN) jets and Gamma Ray Bursts.
 The secondary products of photomeson interactions decay, and
lead eventually to the emission of neutrino and
gamma-ray fluxes from the source, which may be observable. This has
triggered several authors (e.g. Stecker 1973 \cite{Ste68,Ste73}, Berezinsky \&
Gazizov 1993 \cite{BG93}, 
Mannheim \& Schlickeiser 1994 \cite{MS94}) to study photopion
production in more detail. Collider measurements indicate a complicated
structure of the cross section, especially in the astrophysically important
lower energy range. In order to overcome this problem, either simplified
approximations for the cross section, e.g. the so-called
$\Delta$--approximation (Stecker 1973 \cite{Ste73}, 
Gaisser et al. 1995 \cite{GHZ95}), or Stecker's isobar model
\cite{Ste68}, have been used. Berezinsky \& Gazizov \cite{BG93} used
interpolated cross section measurements and implemented them into numerical
codes to derive neutrino production spectra.  Recently, it has been shown
that in realistic photon fields photon--proton collisions may happen at low
{\it and} high center--of--mass (CM) energies (M\"ucke et al
1999a~\cite{Metal99a}), making a
more realistic treatment of the pion production process necessary. For
example, hadronic interactions of ultrarelativistic protons in flat ambient
photon spectra, e.g. in Gamma Ray Burst radiation fields and the synchrotron
radiation field in $\gamma$-ray loud AGN, are roughly equally likely to occur
in the resonance and the multiparticle production regions.

In this paper we discuss photopion production in typical GRB and AGN jet
radiation fields.  We discuss microphysical quantities relevant for the
photohadronic production of gamma-rays, neutrinos, and neutrons (Section 2),
by using our newly developed Monte Carlo event generator SOPHIA (presented
also in these proceedings (M\"ucke et al 1999b \cite{Metal99b})), and compare 
them with results
from a widely used approximation (Section 3). As a qualitatively new effect we
also discuss the photoproduction of anti-baryons (Section 4). The relevance
of our results is briefly discussed in the context of neutrino and cosmic ray
production (Section 5).

\section{Relevant microphysical quantities}

One of the major motivations for discussing photohadronic interactions in
astrophysical sources like AGN jets and GRB is the prediction of observable
neutrino fluxes from these systems. The total flux of neutrinos
for a specific source model can be related to the predicted fluxes
of gamma rays and cosmic rays, which in turn can be compared with 
current observations. Neutrino fluxes from hadronic AGN jet models are
not expected to be observable as point sources by current neutrino telescopes,
and therefore contribute to the extragalactic neutrino background.
 Hence, estimates for diffuse neutrino fluxes originating from blazar
 jets can be derived from measurements of the diffuse extragalactic
 gamma ray background (EGRB) (see e.g. Mannheim 1995~\cite{Man95}) if
 the entire EGRB is produced photohadronically. Upper limits can be set
 if only part of the EGRB is due to unresolved blazars (see
 Chiang \& Mukherjee 1997~\cite{CM97}, M\"ucke \& Pohl 1998~\cite{MP98}). 

Of particular relevance is here the production of gamma rays. In
photohadronic sources, the opacity for the primary gamma rays emerging from
the decay of neutral pions must be large, since otherwise the efficiency
for the photohadronic interactions themselves would be too low to produce
observable fluxes. The interactions of primary photons (and electrons)
 cause
 electromagnetic cascades which reprocess the leptonic
power to lower photon energies, and photons are eventually emitted in the
range $<100$ GeV (Mannheim 1993 \cite{Man93}). 

The exact relation between the gamma ray and neutrino fluxes is determined
by the microphysics of photomeson production, i.e., by the photon-to-neutrino
total energy ratio per interaction, ${\cal E}_{\gamma}/{\cal E}_{\nu} = \sum
E_{\gamma}/\sum E_{\nu}$. The $\Delta$--approximation, which has been
often used, gives ${\cal E}_{\gamma}/{\cal E}_{\nu} = 3$. This ratio has
been slightly modified by considering other photon
producing processes, like Bethe-Heitler pair production off the energetic
protons (Mannheim 1993 \cite{Man93}).

 Another way to constrain neutrino fluxes is a comparison
 with the cosmic ray spectrum. This is motivated by the fact
 that neutrino producing interactions also cause an isospin-flip
 of the proton into a neutron, which is not magnetically confined
 and can be ejected from the accelerator. In contrast, protons
 need to be confined in order to be accelerated, and their
 escape from sources  involving relativistic flows, like AGN 
 jets or GRB, is strongly affected by adiabatic losses. 
 Neutron ejection can also be suppressed if the opacity
 for $n\gamma$ interactions exceeds unity. It is possible
 however to relate the $n\gamma$ opacity to the $\gamma\gamma$
 opacity of very high energy photons, and one can show
 that sources transparent to very high energy (VHE) gamma rays
 are also transparent to neutrons at most energies~\cite{MPR99}. 
 Then, regardless of the proton confinement, the {\em minimum}
 cosmic ray ejection is given by the produced neutron flux,
 which allows an upper limit on the possible neutrino flux
 (Waxman \& Bahcall 1999 \cite{WB99}). It has been pointed
 out by Mannheim et al. \cite{MPR99}, that the complicated
 propagation properties of cosmic rays make it difficult
 to apply a model independent cosmic ray bound on the neutrino flux.
The relation of the cosmic ray flux to neutrino flux is again determined 
by a microphysical parameter, namely the neutrino-to-neutron total energy 
relation ${\cal E}_{\nu}/{\cal E}_n = \sum E_{\nu}/\sum E_{n}$, 
which is predicted $\approx 0.19$ in the $\Delta$--approximation.

Other kinematical quantities of interest are the
inelasticity of the proton, $\Delta E_p/E_p$, which determines the proton
energy loss time scale due to photohadronic interactions, and the average
fractional neutrino energy per interaction, $\langle E_\nu\rangle/E_p$, which
determines the maximum neutrino energy if the maximum proton energy is
determined by the acceleration process. In the same way, we may also
consider the fractional neutron energy per interaction.    

\section{Cosmic ray, gamma ray and neutrino production}

In the typical scenario of hadronic interactions in AGN jet and GRB 
radiation fields, relativistic nucleons with energy $E_p \gg \epsilon$ 
are assumed to interact in an isotropic radiation field in the comoving 
frame of the relativistic plasma flow. Because of the strong magnetic 
fields required to accelerate protons, photoproduced electrons and 
positrons can be assumed to have 100\% radiation efficiency, and transform 
their energy completely into photons through synchrotron emission. The 
total and average energy of the  produced neutrinos includes
$\nu_e$ and $\nu_\mu$ and their antiparticles. The $\nu_\mu/\bar \nu_\mu$
contribution, possibly detectable in current or planned underwater/-ice 
neutrino experiments, is roughly $2/3$ of the total neutrino flux. 
We assume that protons are magnetically confined, and count only
photohadronically produced neutrons as ejected cosmic rays. 

\subsection{Blazar/AGN jets}

There exists a large variety of models to explain the observed $\gamma$-ray
emission from blazars (= flat spectrum radio quasars and BL~Lac
objects). The leptonic models, which assume inverse Compton scattering
of low energy photons up to gamma ray energies, currently dominate the
thinking of the scientific community.  Alternatively, photopion production
has been proposed to be the origin for the observed $\gamma$-ray flux
(Mannheim 1993 \cite{Man93}). A clear distinction between the two models 
is the production of
neutrinos, which is negligible in the leptonic models, but may be equally
important as the photon production in the hadronic models, and has been
predicted to cause observable fluxes in large underwater/-ice neutrino
detectors.

Gamma ray and neutrino production in hadronic blazar models occurs through
photopion production (and subsequent cascading) of relativistic protons in
either external (accretion disk or the IR radiation field from a molecular
torus) radiation fields or the synchrotron radiation field produced by
electrons which are shock accelerated together with the protons.  Here, we
confine our discussion to the latter model. For simplicity, we consider
TeV-blazars only, which are characterized by a low energy spectrum extending
to X-ray energies.  The synchrotron emission from blazar jets at low energies
can be well explained by a superposition of several self-absorbed synchrotron
components (e.g. Cotton et al 1980 \cite{CWS80}, 
Shaffer \& Marscher 1979 \cite{SM79}) leading to a
flat target spectrum.  The {\it{local}} synchrotron radiation spectrum
follows a power law with photon index $\alpha=1.5$ ($n(\epsilon) \propto
\epsilon^{-\alpha}$) up to a break energy (see e.g. Mannheim 
1993 \cite{Man93}).  Above
the break energy $\epsilon_b \approx 10^{-4}$eV 
(see e.g. Rachen, \& M{\'e}sz{\'a}ros \cite{RM98}) it is loss
dominated to become $n(\epsilon) \propto \epsilon^{-2}$. The synchrotron
emission in TeV-blazars like Mkn~421 and Mkn~501 is observed to continue up
to 1-10 keV. Thus, for our application we approximate the typical target
photon spectrum in the jet frame (assuming a Doppler factor of D=10) by
\begin{eqnarray}
n(\epsilon) & \propto &  \epsilon^{-3/2} \qquad \mbox{for} \qquad
 10^{-5}\rm{eV} \leq \epsilon \leq 10^{-4}\rm{eV}\nonumber\\
n(\epsilon) & \propto & \epsilon^{-2\phantom{/2}} \qquad  \mbox{for} \qquad
 10^{-4}\rm{eV} \leq \epsilon \leq 10^{3}\rm{eV}
\end{eqnarray}
For the maximum proton energy we use $E_{p,\rm max} = \gamma_p m_p c^2 = 
10^{10}$~GeV in the rest frame of the jet, corresponding roughly to the
highest observed cosmic ray energies after Doppler boosting. 
This is also consistent in order of magnitude with the
values derived by equating the acceleration time of the proton 
with its total loss time due to adiabatic, synchrotron and photohadronic 
cooling (see Rachen \& M\'esz\'aros \cite{RM98} for a more detailed 
discussion). Due to the threshold
condition for photopion production of protons with Lorentz factor $\gamma_p$
interacting with photons of energy $\epsilon$
\begin{equation}
\gamma_p > \frac{m_\pi c^2}{2\epsilon} \left(1+\frac{m_\pi}{2m_p}\right )
\end{equation}
only photons from the steep part of the target radiation field interact
photohadronically with protons of $\gamma_p \leq 10^{10}$.

\begin{figure}
\vbox{\psfig{figure=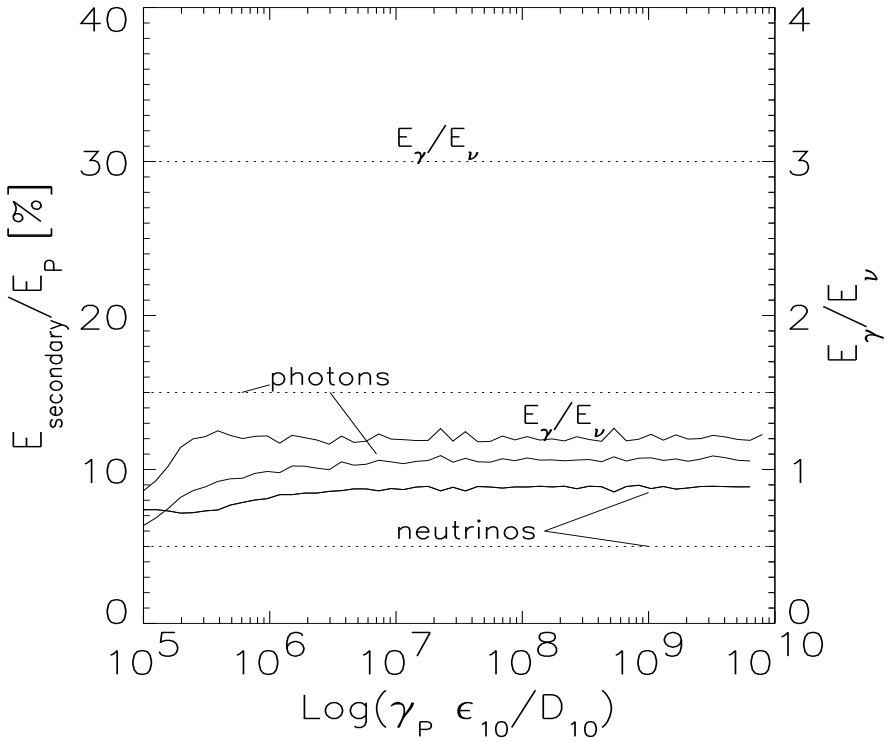,height=10cm,width=12cm,clip=}}
\caption[fig1]{
\label{content-AGN}
The total photon and $\nu$-power (normalized to
the proton input energy $E_p$), and their ratio emerging from a
source due to photopion production of protons of energy $E_p$ in a typical
TeV-blazar ambient radiation field The ${\cal E}_{\gamma}/{\cal
E}_{\nu}$--ratio is approximately unity, independent of proton input energy,
which can be contrasted to the $\Delta$-approximation (dotted lines).
Solid lines denote the results from the SOPHIA simulations.}
\end{figure}

Fig.~\ref{content-AGN} shows ${\cal E}_\gamma/{\cal E}_\nu$
from SOPHIA simulations for protons of energy $E_p$ interacting 
in the synchrotron field of blazar jets. 
The prominent resonance/threshold region dominates the
interaction, independently of input proton energy. 
Consequently, the total fractional photon energy ${\cal E}_\gamma/E_p 
\approx 0.1$, independent of $E_p$, and an overall photon-to-neutrino 
ratio of ${\cal E}_{\gamma}/{\cal E}_{\nu}\approx 1.2$.
We note that this result differs by a factor $\approx 3$ from the 
prediction of the $\Delta$-approximation, although the steep spectrum 
emphasizes the threshold region with the dominant $\Delta(1232)$ resonance. 
This can be understood in view of the strong contribution of non-resonant 
direct $\pi^+$ production at threshold, and the contribution of other 
resonances with different isospin in the vicinity of the $\Delta(1232)$ 
resonance (see M\"ucke et al., these proceedings \cite{Metal99b}).

\begin{figure}
\vbox{\psfig{figure=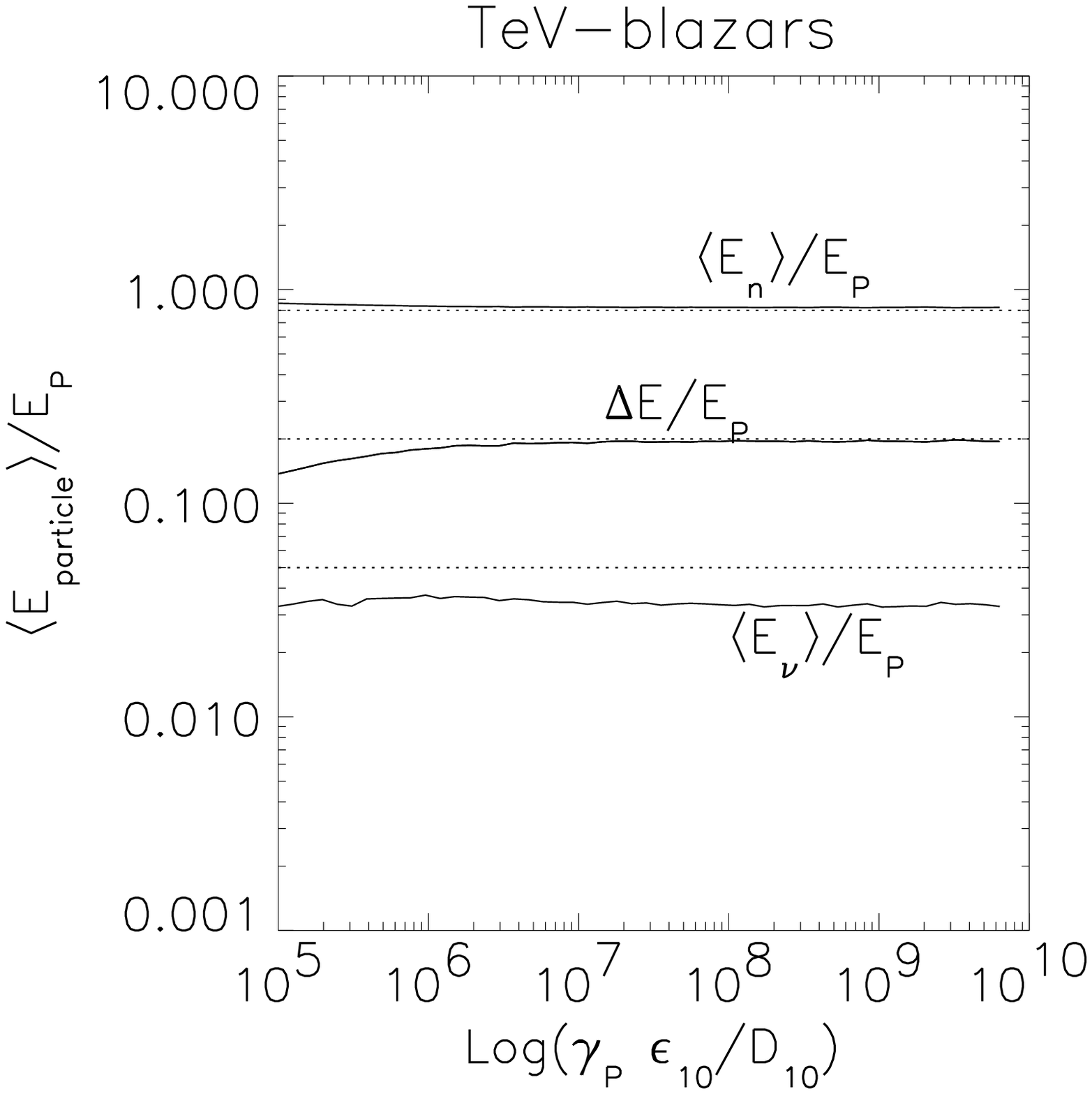,height=10cm,width=12cm,clip=}}
\caption[fig2]{
\label{elanu-PL}
Average neutron (upper curves) and neutrino energy (lower curves) with respect
to the proton input energy $E_p$, and the mean proton inelasticity
 due to photon--proton pion production of protons with energy $E_p$ in
a typical TeV-blazar ambient radiation field.  The dotted lines represent the
respective values from the $\Delta$-approximation while
solid lines denote the results from the SOPHIA simulations.}
\end{figure}

\begin{figure}
\vbox{\psfig{figure=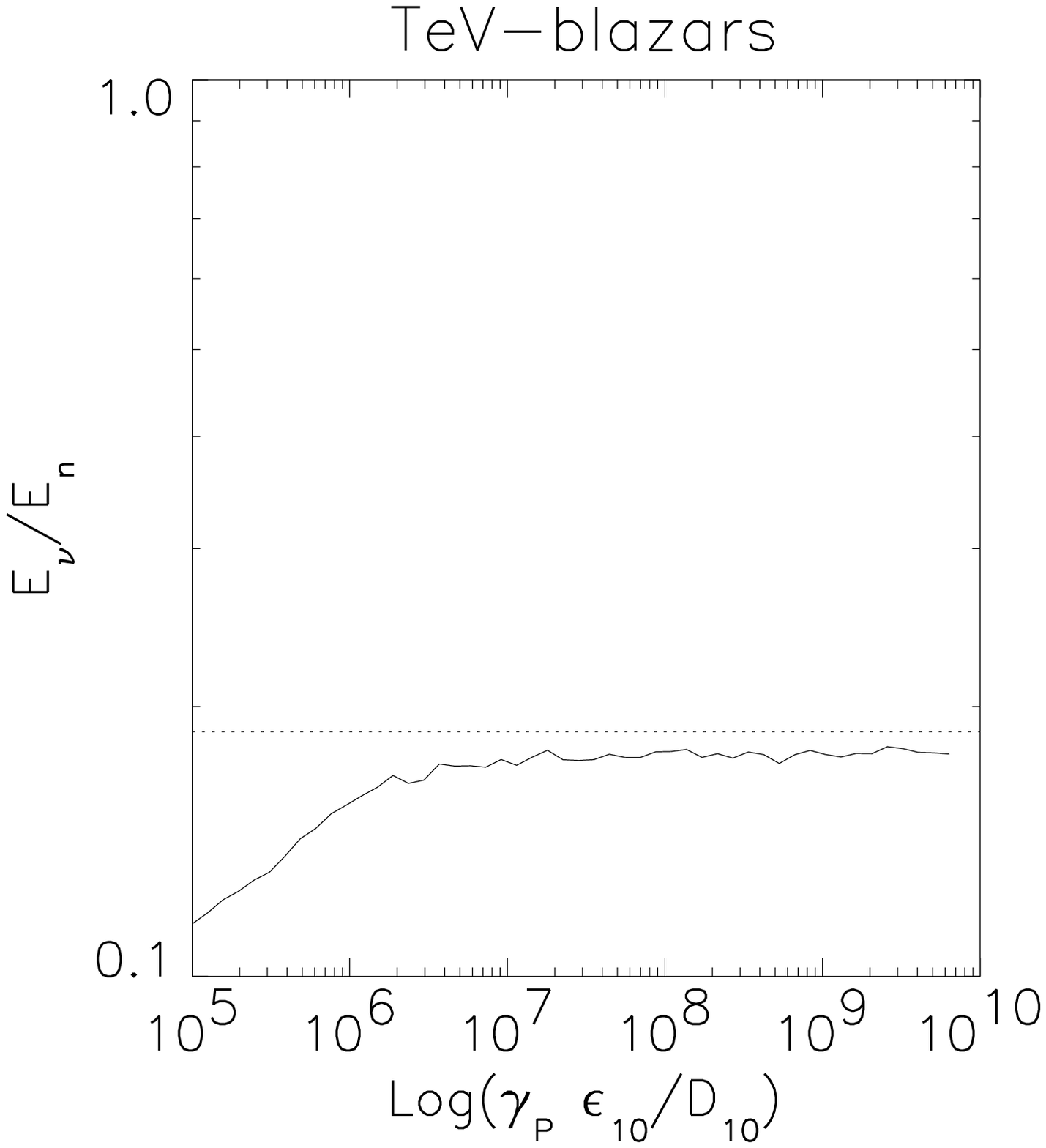,height=10cm,width=12cm,clip=}}
\caption[fig3]{
\label{n+nu-PL}
The ratio of total $\nu$-energy to neutron energy emerging from a TeV-blazar
jet due to photopion production of a protons with energy $E_p$.  The
$\Delta$--approximation (dotted line) gives an adequate description of the
$\nu$-to-neutron ratio.}
\end{figure}

In contrast, kinematical quantities like inelasticity and fractional secondary
particle energy are well reproduced by the $\Delta$-approximation (see
Fig.~\ref{elanu-PL}), since these quantities vary only slowly with energy. 
For example, the proton inelasticity is in general logarithmically rising
with increasing nucleon input energy in the resonance region
(M\"ucke et al 1999b \cite{Metal99b}). 
Thus, when interactions
occur near the $\Delta(1232)$-resonance like in this case, the
$\Delta$--approximation gives a good description. The same applies for
the total neutrino-to-neutron energy ratio. Here, we find 
${\cal E}_{\nu}/{\cal E}_n\approx 0.18$ (see Fig.~\ref{n+nu-PL}).

\subsection{Gamma Ray Bursts}

The cosmological Gamma Ray Burst fireball model has been very successful in
explaining the observed temporal evolution of the afterglow photon
spectra (see Piran 1998~\cite{P98} and references therein). 
In this model a relativistically expanding fireball transforms
most of the explosion energy into the kinetic energy of baryons in a
relativistic blast wave with bulk Lorentz factor of 100 -- 300.
This energy is reconverted into radiation at shocks, produced either in
collisions between different shells ejected from the central source (internal
shock scenario), or through deceleration of the blast wave when it sweeps up
the external medium (external shock scenario). 
The former process is thought to be responsible for the GRB itself, 
while the latter produces
the afterglow. In this scenario, the observed photon radiation is explained
as synchrotron radiation from the accelerated electrons. In the comoving
shock frame this target radiation field for photopion production of the
relativistic protons may be approximated by a broken power law:
\begin{eqnarray}
n(\epsilon) & \propto &  \epsilon^{-2/3} \qquad \mbox{for} \qquad
 10^{-3}\rm{eV} \leq \epsilon \leq 10^{3}\rm{eV}\nonumber\\
n(\epsilon) & \propto & \epsilon^{-2\phantom{/2}} \qquad  \mbox{for} \qquad
 10^{3}\rm{eV} \leq \epsilon \leq 10^{5}\rm{eV}
\end{eqnarray}
Waxman \cite{Wax95} and Vietri \cite{Vie95} suggested that if a significant
fraction of the observed GRB power is transformed into ultrahigh energy
cosmic rays (UHECRs), GRB may well be the source for all the observed UHECRs.
In fact, it has been shown that comoving proton Lorentz factors of up to
$\sim 10^9$ can be reached in GRB shells, which are boosted in the observers
frame to $3 \cdot 10^{20}$ eV (\cite{Wax95},\cite{RM98}). 
This scenario gives also rise to fluxes of
very high energy neutrinos ($E_\nu > 100$ TeV) correlated with gamma ray
bursts, which could be detected in a km$^3$ neutrino observatory
because the exact temporal and directional information reduces the 
background to virtually zero (Waxman \& Bahcall 1997 \cite{WB97}).
 The neutrino emission, however, 
is suppressed at
the highest energies because of the adiabatic and synchrotron 
losses of pions and muons prior to their 
decay (Waxman \& Bahcall 1997 \cite{WB97},
Rachen \& M{\'e}sz{\'a}ros 1998 \cite{RM98}, 
Waxman \& Bahcall 1999 \cite{WB99}).

\begin{figure}
\vbox{\psfig{figure=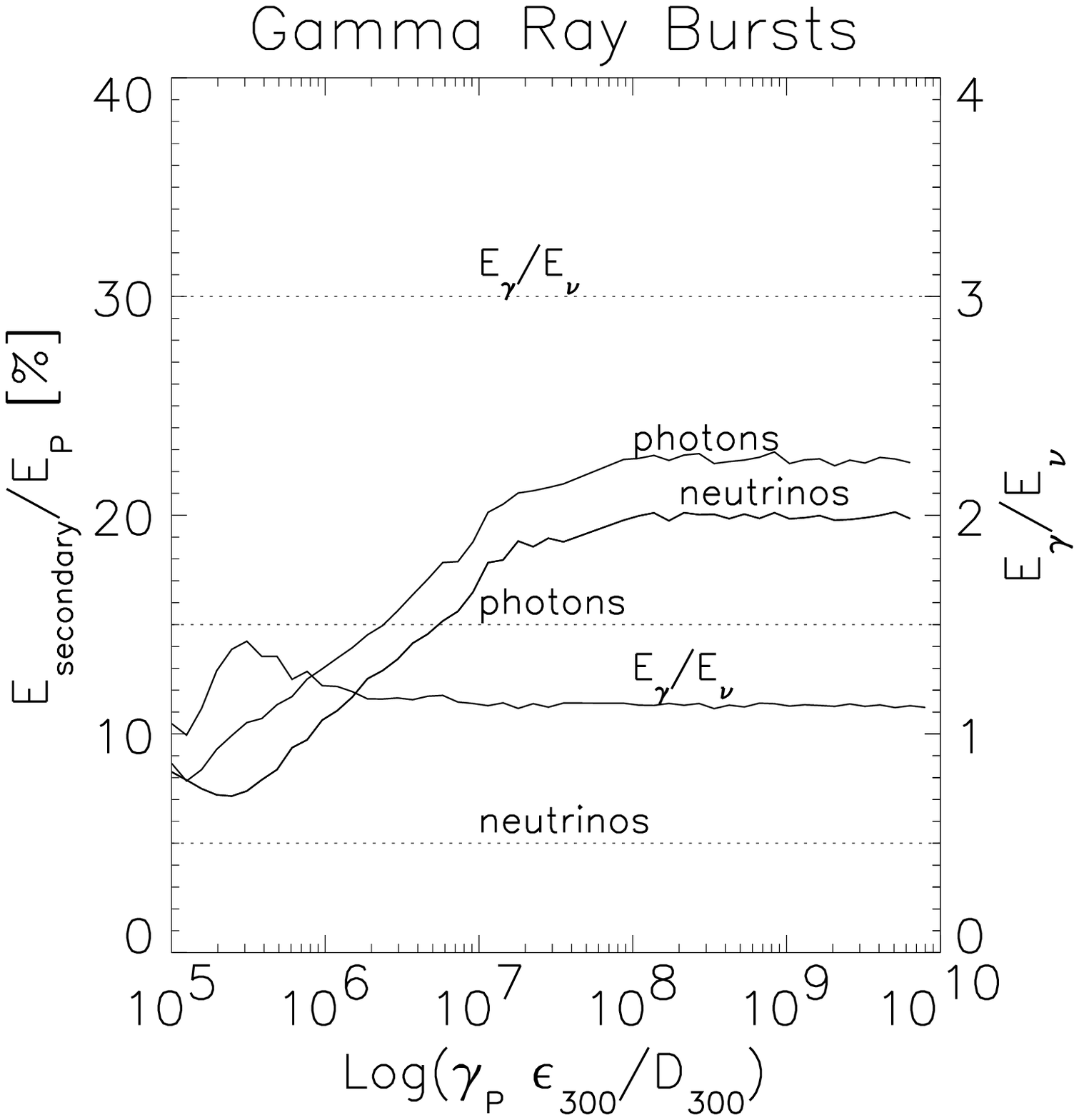,height=10cm,width=12cm,clip=}}
\caption[fig1]{
\label{content-GRB}
The total power in photons and neutrinos (normalized to
the proton input energy $E_p$), and their ratio due to photomeson
production in a typical GRB ambient radiation field. The ${\cal
E}_{\gamma}/{\cal E}_{\nu}$--ratio is nearly always unity, independent of
proton input energy, which can be contrasted to the $\Delta$-approximation
(dotted lines). Solid lines denote the results from the SOPHIA simulations.  }
\end{figure}

In the highly energetic GRB photon field mostly photons from the flat part of
the ambient radiation field interact photohadronically. For incident protons
with $\gamma_p > 10^7$ interactions predominantly occur in the
multiparticle region of the cross section, while for $\gamma_p \leq 10^7$ the
resonance/threshold region determines the secondary particle production.
This leads to an increase of photon and neutrino production by roughly a
factor of 2 with ${\cal E}_{\gamma} \approx {\cal E}_{\nu}$ for
protons with $\gamma_p > 10^7$ in comparison with protons with $\gamma_p
\approx 10^5$ (see Fig.~\ref{content-GRB}).  The photon-to-neutrino total
energy ratio seems to be fairly robust ${\cal E}_{\gamma} \approx {\cal
E}_{\nu} \approx 1$ at all proton energies, except for a slight deviation at
$\gamma_p \approx 3 \cdot 10^5$ where photons 
from the flat part of the target photon spectrum
interact mainly via the $\Delta(1232)$-resonance. Also here, this result is 
about a factor of 3 different from the $\Delta$--approximation.

\begin{figure}
\vbox{\psfig{figure=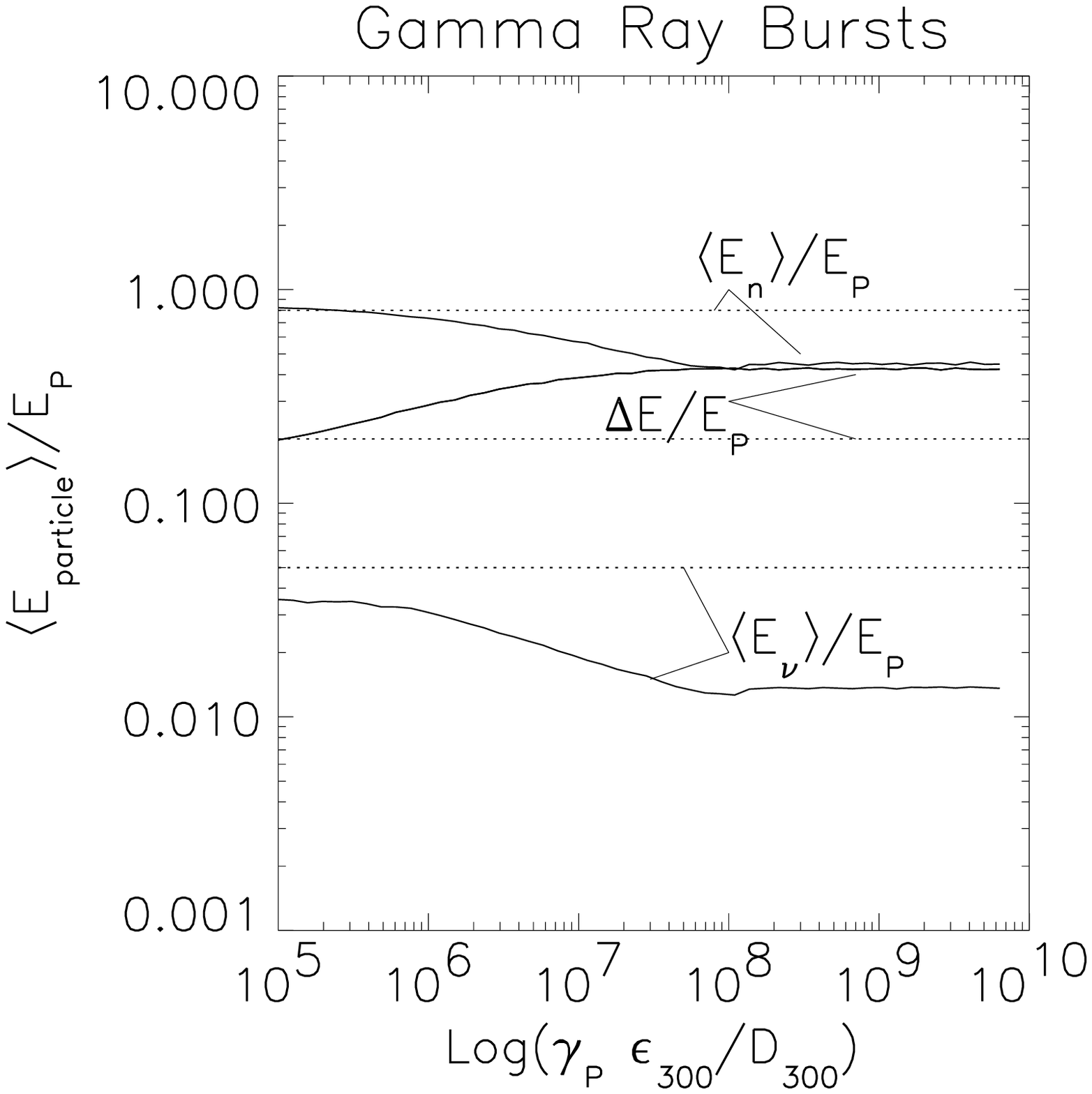,height=10cm,width=12cm,clip=}}
\caption[fig5]{
\label{elanu-GRB}
Average neutron and neutrino energy with respect
to the proton input energy $E_p$, and the mean proton inelasticity
 due to photon--proton pion production in a typical GRB radiation
field. The dotted lines represent the respective values from the
$\Delta$-approximation which are only met for interactions near threshold,
e.g. at low 
input energies $\gamma_p < 10^6$. Solid lines denote the results from 
the SOPHIA simulations. }
\end{figure}

\begin{figure}
\vbox{\psfig{figure=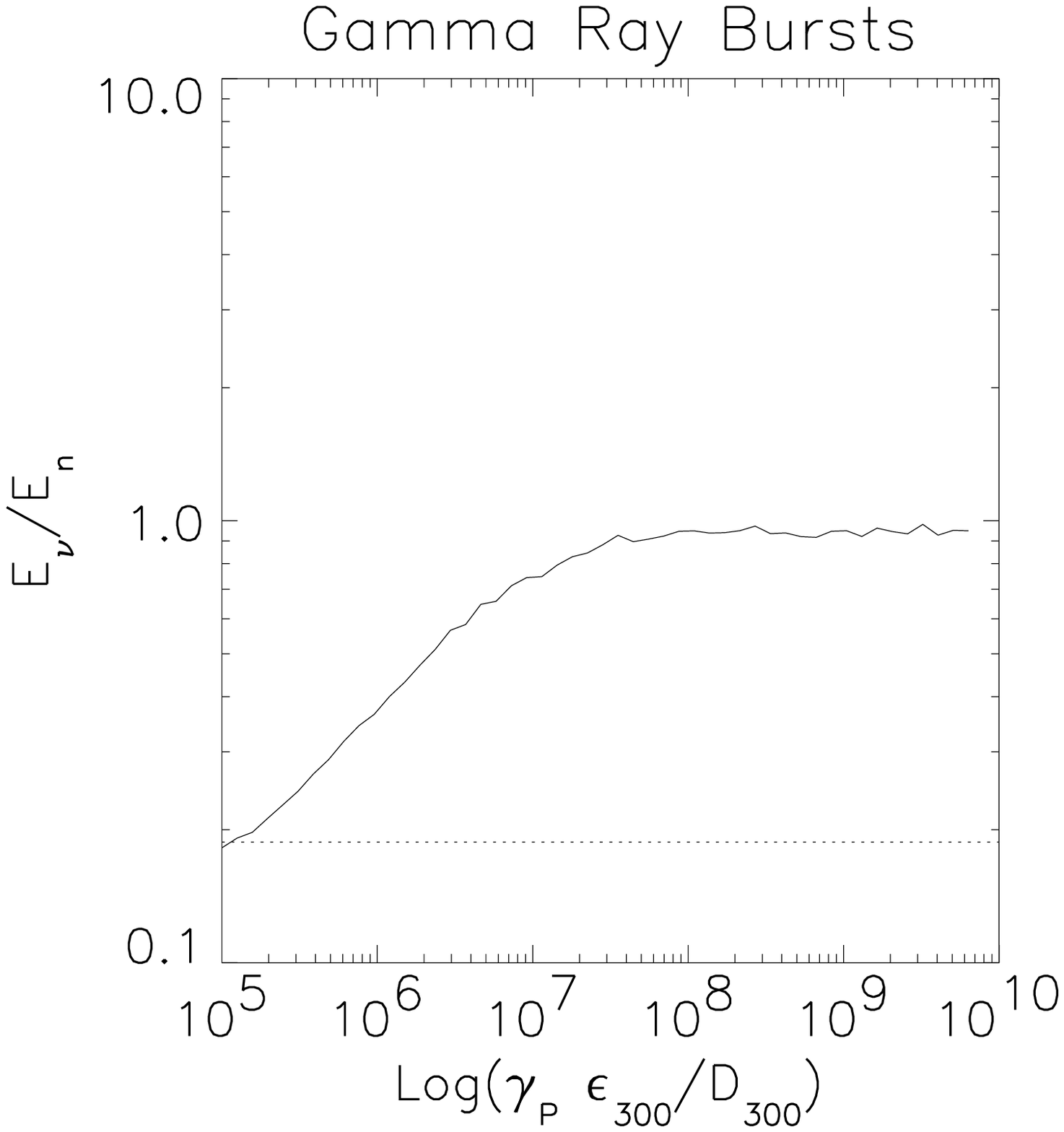,height=10cm,width=12cm,clip=}}
\caption[fig6]{
\label{n+nu-GRB}
The total $\nu$-energy to neutron energy emerging from a source due to
photopion production in a typical GRB photon field.  With increasing input
proton energy the total neutrino power per produced cosmic ray is
underestimated by a factor $< 5$ in comparison to the $\Delta$--approximation
(dotted line).  }
\end{figure}

The average number of neutrinos produced per interaction can increase
 with CM energy by up to an order of magnitude for the proton energy
 range relevant in GRB (see M\"ucke et al 1999b \cite{Metal99b}). 
Because of the high multiplicity of the secondaries, the mean energy of
the neutrinos produced in the multipion region (i.e. for $\gamma_p > 10^7$)
is $\approx 1\%$ of the proton input energy (see Fig.~\ref{elanu-GRB}). 
At lower energies it reaches $\langle E_\nu\rangle/E_p \approx 0.04$, 
approximately in agreement with the
$\Delta$--approximation. Therefore, the ejection of ultra-high energy
neutrinos ($E_\nu >10^{19}$ eV), as recently proposed by Vietri \cite{Vie98}
(but note also \cite{WB99}) requires observer frame proton energies
of $10^{21}$ eV.
Such high proton energies are not expected from shock acceleration scenarios 
in GRB shells. In the general case, however, the synchrotron losses 
of charged pions and muons in the highly magnetic ($B \sim 10^3 - 10^6$G) 
GRB environment  and their adiabatic deceleration determine the high 
energy end of the observable neutrino flux from GRB 
(see Rachen \& M{\'e}sz{\'a}ros 1998 \cite{RM98}).

 If protons are magnetically confined at the source, the proton--neutron
 conversion probability determines the production of cosmic rays in
 photohadronic sources. At high energies it decreases from
 $\approx 0.5$ to $0.3$ (see M\"ucke et al, these proceedings~\cite{Metal99b}),
 and the average fractional energy of neutrons decreases to $<0.5$ of
 the proton energy. Together with the increasing power 
 going into secondary particles, this leads to a neutrino-to-neutron 
 ratio ${\cal E}_{\nu}/{\cal E}_n \approx 1$, about a factor of $5$ 
 above the $\Delta$--approximation. This increase of the neutrino
 power relative to the cosmic ray power of GRB (if protons are confined)
 is, however, masked by the secondary particle losses
 (see Rachen \& M{\'e}sz{\'a}ros 1998 \cite{RM98}, 
 Waxman \& Bahcall 1999 \cite{WB99}).
 At lower energies
 the neutrino-to-neutron ratio approaches the value expected
 in the $\Delta$-approximation (see Fig.~\ref{n+nu-GRB}).

\section{High energy anti-baryon production}

A qualitatively new effect which can be investigated with SOPHIA is the
production of secondary baryon/anti-baryon pairs in high energy photohadronic
interactions. This requires a detailed simulation of QCD string
hadronization, which is implemented in SOPHIA.  The
theoretical threshold of this process is $\sqrt{s} = 3m_p c^2$, 
corresponding to a photon energy in the proton rest frame 
$\epsilon' = 4 m_p c^2$. For $\epsilon' > 2$ TeV, about $40\%$ of 
all events contain antinucleons. This raises the interesting question 
about the contribution of antiprotons to the
cosmic ray flux from extragalactic photopion production sources.

 Due to the threshold condition, this process can only affect the
 antimatter to matter ratio at extremely high energy.
Assuming the minimal cosmic ray ejection
hypothesis, (only neutral baryons can leave the source and become cosmic rays)
the relevant quantities are the anti-neutron to neutron ratio 
${\cal E}_{\bar n}/{\cal E}_n$ and the corresponding
multiplicity ratio $N_{\bar n}/N_n$.

As noted above, cosmic rays with energy $\gamma_p > 10^7$ 
accelerated in GRBs tend to interact with photons from the 
flat part of the ambient radiation fields, in the multiparticle 
production regime. While the total energy dissipated into 
photons, neutrinos and neutrons is roughly the same (${\cal
E}_{\gamma} \approx {\cal E}_{\nu} \approx {\cal E}_{n} \approx 0.2 E_p$) 
in this high energy part of the cross section, the antineutron 
production is only 1/40 of the neutron production. The neutrons 
carry on average 50\% of the input energy (see Fig.~\ref{elanu-GRB})
with a neutron multiplicity of roughly 0.4.
Antineutrons have mean energies of $\simeq 0.1 E_p$ with a
multiplicity of approximately 0.05. 
Antineutrons and neutrons of the same energy are produced
by protons of different energy.
For example, we need an input proton energy of $10^{10}$GeV 
for the production of a antineutron of energy $10^9$GeV while 
for a neutron of the same energy the proton input energy must 
be $\approx 2\cdot 10^9$GeV.
One has to weight the ratio of the resulting neutron 
and antineutron multiplicities, and this is determined 
by the input proton spectrum.  
For this purpose we use a $E^{-2}$ differential spectrum
as a typical equilibrium particle spectrum for GRB.
For photon--proton interactions at high
CM energies, as is the case for GRBs, we therefore expect
antineutron-to-neutron multiplicities of the order 0.01.

Since photon--proton interactions in TeV-blazar ambient 
photon fields take part predominantly at threshold, 
these objects are not expected to be strong sources of 
antinucleons.  SOPHIA simulations show that the total energy
for antineutron production does not exceed $10^{-4}$ of 
the proton input energy with the average antineutron carrying 
roughly 1/10 of $E_p$. Neutrons, on the other hand, take up 
about 20\% of the input energy with a mean particle
energy of $0.8 E_p$, as expected from the $\Delta$--approximation. 
Weighting again with the typical equilibrium differential 
particle spectrum expected in blazar jet environments, which follows a $E^{-1}$ power law,
we find that the $\bar n/n$--ratio does not exceed $10^{-3}$.

\begin{figure}
\vbox{\psfig{figure=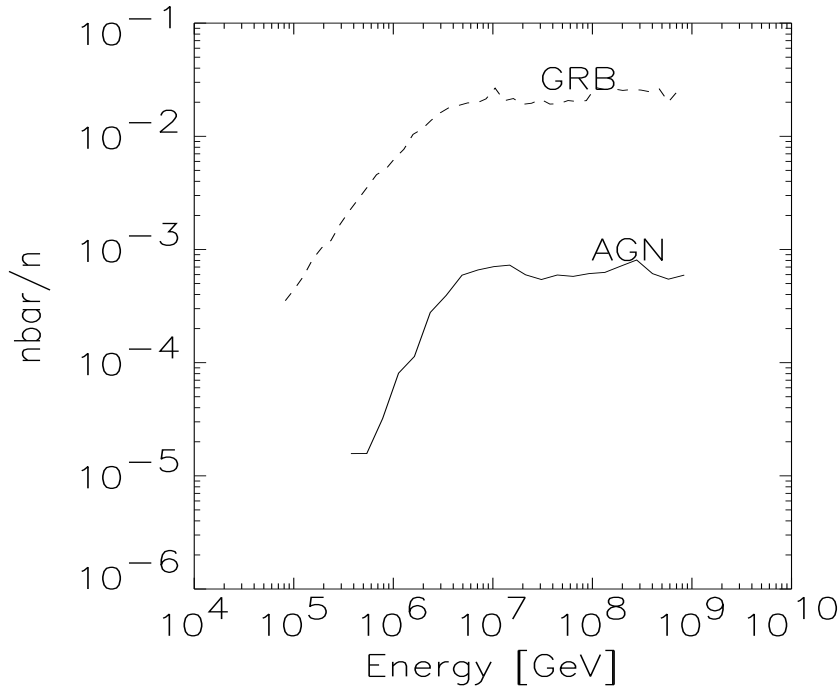,height=10cm,width=12cm,clip=}}
\caption[fig8]{
\label{ppbar}
Predicted antineutron-to-neutron ratio at production in the shock frame of
Gamma Ray Burst (GRB) and in TeV-blazar jets. GRBs predict a
antibaryon-to-baryon production ratio, which approaches 0.01 above
$10^{16}$eV, while TeV-blazars may produce antibaryons on a
level of about one order of magnitude lower.}
\end{figure}

The $\bar n/n$--ratios resulting from SOPHIA in GRB and
TeV-blazar radiation fields as a function of proton energy 
are shown in Fig.~\ref{ppbar}. After $\beta$-decay, the 
ejected $\bar n/n$ contribute to the overall $\bar p/p$ flux. 
The $\bar n/n$ ratio derived here has to be
regarded as an upper limit to the observable $\bar p/p$ ratio, 
since direct proton injection may contribute to the cosmic ray 
flux without increasing the anti-proton population.

\section{Discussion and outlook}

Our simulation results obtained with the new photomeson production event
generator SOPHIA \cite{Metal99b} demonstrate that the widely used
$\Delta$-approximation for photohadronic interactions has only a 
limited applicability to many important astrophysical 
applications, namely the secondary gamma ray, neutrino and 
neutron production in radiation fields of AGN jets and GRB. 
For AGN jets, where interactions at low center--of--mass energy 
dominate, the major kinematical quantities, like proton 
inelasticity, average fractional neutrino energy per interaction, 
and the neutrino-to-neutron energy ratio are found to be well
predicted by the $\Delta$-approximation. For interactions of 
 ultra--high energy cosmic rays in GRB shells, however, we find 
 deviations up to an order of magnitude for some of these quantities. 

As a robust result for both kinds of sources we find that the average
total energy channeled into neutrinos, ${\cal E}_\nu$, and gamma-rays, 
${\cal E}_\gamma$ are approximately equal. By comparison, 
the $\Delta$--approximation predicts ${\cal E}_\gamma/{\cal E}_\nu 
\approx 3$. This relation has been widely used to normalize 
expected neutrino fluxes from AGN jets to their photon flux. 
As a consequence of this deviation from the $\Delta$--approximation 
found in our SOPHIA results,  the expected neutrino fluxes from such 
AGN models would increase by a factor of $3$ assuming that the 
observed gamma ray emission is entirely of photohadronic origin.
The lower gamma-ray-to-neutrino ratio seen in our SOPHIA results, 
together with the neutrino-to-neutron ratio in agreement with the 
$\Delta$--approximation, implies also that AGN neutrino models, 
which were initially designed to comply with the limitation set by 
the cosmic ray data in a straight line propagation scenario (Mannheim,
1995 \cite{Man95}, Model A), can no longer produce a large fraction 
of the 100 MeV gamma ray background as initially assumed. This may comply
with the recent finding of Chiang \& Mukherjee (1997 ~\cite{CM97}) that
only part of the measured EGRB may be due to unresolved blazars.
However, it has been pointed out by Mannheim et al.\ \cite{MPR99} that an 
increased and so far unexpected contribution from sources with 
maximum cosmic ray energies below $10^{19}$ eV can evade the problem 
without being in conflict with the cosmic ray data. 
 Moreover, it is possible that the optical depth of blazars for
 ultra-high energy neutrons is larger than unity, which allows
 a higher diffuse neutrino and gamma ray flux for a given cosmic ray flux.

For the derivation of the upper bound of the neutrino emission, 
the neutrino-to-neutron energy ratio reflects the microphysics
of the photohadronic interactions. The value
${\cal E}_\nu/{\cal E}_n\approx 0.19$, expected from the
$\Delta$-approximation and used in Waxman and Bahcall \cite{WB99}, 
was confirmed for AGN jets in our simulations. For GRB, however, 
we find  a ratio of about $1$. This implies that the cosmic ray
upper bound for GRB neutrinos has to be increased by a factor of 
$5$ at high energies. In practice this theoretical bound 
is unlikely to be reached by GRB neutrinos, since their emission 
is suppressed at high energies by secondary particle (pion 
and muon) cooling in the hadronic cascade \cite{WB97,RM98}. 
Models which try to evade such losses by shifting the acceleration 
region into the outer shock and the afterglow of the GRB, 
as suggested by Vietri (\cite{Vie98}; note also \cite{WB99}), 
 can be shown not to be efficient enough 
 to reach such fluxes for reasonable energetics \cite{RM98}. 

The average neutrino energy due to photopion production 
in GRB spectra at very high proton energies is up to an 
order of magnitude below the value expected from the 
$\Delta$-approximation. This has severe consequences
for the prospects of GRB--correlated neutrino events 
above $10^{19}$ eV. The fact that the mean proton energy 
loss rises with increasing proton energy up to $\Delta E/E \simeq 0.5$ 
(compared to the $\Delta$--approximation value $\Delta E/E = 0.2$) 
may add to the problem, since it leads to lower proton 
energies in photohadronically limited acceleration scenarios.

A qualitatively new feature which can be explored with 
SOPHIA is the photoproduction of antibaryons \cite{Metal99b}. 
Our prediction of the maximal antiproton contribution from 
GRB of about $2\%$, reached for energies above $5\cdot 10^{17}$ eV, 
is several orders of magnitude above the background expected at
this energy from cosmic ray-nucleon collisions in the galactic disk. 
The latter is measured at energies $\sim 10$ GeV as approximately 
$10^{-3}$, and expected to decrease with $E_p^{-[0.3{-}0.6]}$ following 
the leaky box model. Since the expectation of a significant increase 
of the $\bar p$ contribution at high energies is unique for GRB, 
this could provide an independent test whether GRB are indeed the 
dominant sources of cosmic rays. Unfortunately, there is presently 
no imaginable way to distinguish between nucleons and anti-nucleons 
at such high energies, where cosmic rays can only be measured in 
air shower experiments. Our result may therefore be
regarded as of rather academic interest, or may be kept in mind 
for presently not anticipated, future detection techniques.

In conclusion, we have demonstrated that the application of 
SOPHIA to astrophysical problems involving the interaction of
energetic cosmic rays in photon backgrounds can (a) improve the
accuracy of the predictions from such models, and (b) open 
the possibility to explore particle physics effects so far 
neglected in astrophysics. 

\section*{Acknowledgments}

The work of AM and RJP is supported by the Australian Research Council.  RE
and TS acknowledge the support by the U.S. Department of Energy under grant
number DE FG02 01 ER 40626. TS is also supported in part by NASA grant
NAG5-7009. The contribution of JPR was supported by NASA NAG5-2857 and 
by the EU-TMR network ``Astro-Plasma Physics'' under contract number
ERBFMRX-CT98-0168.

\section*{References}


\end{document}